\documentclass[%
 reprint,
superscriptaddress,
showpacs,
 amsmath,amssymb,
 aps,
 prl,
]{revtex4-2}
\usepackage{amsfonts}

\usepackage{todonotes}
\usepackage{amsmath}
\usepackage{amsthm}
\usepackage{bm}% bold math
\usepackage{hyperref}

\renewcommand{\phi}{\varphi}
\renewcommand{\rho}{\varrho}

\newcommand{\bp}{{\bm{p}}}
\newcommand{\bP}{{\bm{P}}}
\newcommand{\bx}{{\bm{x}}}
\newcommand{\bX}{{\bm{X}}}
\newcommand{\br}{{\bm{r}}}
\newcommand{\bk}{{\bm{k}}}
\newcommand{\bl}{{\bm{\ell}}}
\newcommand{\bmm}{\bm{m}}

\newcommand{\BB}{\mathrm{BB}}
\newcommand{\IB}{\mathrm{IB}}

\newcommand{\hc}{\mathrm{h.c.}}

\begin{document}

\title{Validity of the Fröhlich model for a mobile impurity in a Bose-Einstein condensate}

\author{Jonas Lampart}
\email{jonas.lampart@u-bourgogne.fr}
\affiliation{Université Bourgogne Europe, CNRS, Laboratoire Interdisciplinaire Carnot de Bourgogne ICB UMR 6303, 21000 Dijon, France}

\author{Arnaud Triay}
\email{triay@math.lmu.de }
\affiliation{Department of Mathematics, LMU Munich, Theresienstraße 39, 80333 Munich, Germany}

\begin{abstract}
We analyze the many-body Hamiltonian describing a mobile impurity immersed in  a Bose-Einstein condensate (BEC). Using exact unitary transformations and rigorous error estimates, we show the validity of the Bogoliubov-Fröhlich Hamiltonian for the Bose polaron in the regime of moderately strong, repulsive interactions with a dilute BEC. Moreover, we calculate analytically the universal logarithmic correction to the ground state energy that arises from an impurity mediated phonon-phonon interaction.
\end{abstract}

\maketitle
\section{Introduction}

The concept of a polaron, a quasi-particle formed through the interaction of an impurity with a medium, is ubiquitous in condensed matter physics~\cite{alexandrov2010, grusdt2016c, grusdt2024}.
Among the many examples, the Bose polaron stands out by the degree of control over the interactions allowed by cold atomic gases.
For this reason, it is seen as a possible platform for the simulation of polaron physics in analogous systems, and in particular allows to access the large-coupling regime~\cite{Joergensen2016, Hu2016, yan2020bose}.
The system also serves a similar role for theory, as a many-body quantum system whose constituents are simple enough that one can understand its complex behavior and  effective degrees of freedom from first principles and has thus attracted considerable attention~\cite{tempere2009,rath2013, li2014, grusdt2015,  christensen2015,grusdt2016d, grusdt2017, drescher2019, ichmoukhamedov2019, levinsen2021,massignan2021, seetharam-21, ichmoukhamedov2022, mostaan2023, seetharam-24}. For the same reason, the rigorous justification of the effective description of fluctuations in dilute Bose gases by Bogoliubov phonons is an active area of mathematical research~\cite{RodSch-09,YauYin-09,GriMacMar-10,GriMacMar-11,Sei-11,LewNamSch-15,LewNamSerSol-15,BocCenSch-17,BocBreCenSch-18,FouSol-20,FouSol-23,NamTri-23,BreSchSch-22, HabHaiNamSeiTri-23,caraci-24}, with recent results also considering impurities~\cite{MySe-20, LaPi22,  LaTr24}.

Polarons are often modeled by the Fröhlich Hamiltonian, in which the impurity interacts with the field of low-energy excitations of the medium by linear coupling. However, the validity of this model for the Bose polaron has been questioned~\cite{christensen2015}.
Given the importance of the Bose polaron as a model system, it is crucial to understand precisely the regime of validity of the Fr\"ohlich Hamiltonian and control the modeling errors.
In this paper, we give conditions under which the Fröhlich Hamiltonian provides a good approximation to the full many-body spectrum of a dilute Bose gas with an impurity. Our arguments are based on the mathematical analysis of \cite{LaTr24} where we prove rigorous error estimates.
%
% In this letter, we show that in the intermediate coupling regime the Fröhlich Hamiltonian provides a good approximation to the full many-body spectrum of a dilute Bose gas with an impurity.
%, and provide rigorous error estimates. 
This allows the model to be used with a great deal of confidence, and certifies the analogies to systems with a similar Hamiltonian.

We find that the total energy of the system has a logarithmic correction to the mean-field terms that is still universal, i.e., depends on the interaction only via the impurity-boson scattering length.
Compared to similar terms in the pure Bose gas~\cite{wu1959, sawada-59, HuPi-59, caraci2023}, it might be more accessible experimentally, by tuning the  impurity-boson interaction.

% Although known since the 50's \cite{bogoliubov1947,LeeHuangYang-57,HuPi-59}, there has been a tremendous mathematical effort to justify that quantum fluctuations in dilute Bose gases are correctly described by Bogoliubov's theory \cite{Sei-11,YauYin-09,BocBreCenSch-18,FouSol-20,FouSol-22,FouGirJunMorOli-24,HabHaiNamSeiTri-23,BasCenOlgPasSch-23}. In the weakly interacting regime, the Bogoliubov-Fr\"ohlich Hamiltonian has been rigorously derived but the logarithmic divergence is absent \cite{MySe-20}. Here we follow \cite{LaTr24} and discuss in detail how the renormalization of the energy and the couplings arises from the underlying many-body quantum system.

We consider an impurity interacting with a dilute Bose gas ($\rho_\mathrm{B}a_\BB^3\ll1$; $\rho_\mathrm{B}$ is the boson density, $a_\BB$ the boson-boson scattering length) at zero temperature.
Both the boson-boson and boson-impurity interactions are assumed to be repulsive.
In this situation, the relevant low-energy excitations are the Bogoliubov phonons.
In order for the Fröhlich Hamiltonian to provide a good approximation, the impurity-boson interaction should
not be so strong as to deplete the Bose-Einstein condensate. On the other hand, it should also be strong enough that the impurity-phonon interaction cannot simply be neglected. This corresponds to the regime in which the dimensionless polaronic coupling, which we express in terms of the scattering lengths and the healing length  $\xi = 1/\sqrt{8\pi \rho_{\BB} a_\BB}$ of the Bose gas as
\begin{equation*}
 \alpha= a_\IB^2 \xi \rho_\mathrm{B}=\frac{a_\IB^2}{8 \pi a_\BB \xi}
\end{equation*}
is of order one~\footnote{Note that defintions of this parameter in~\cite{tempere2009, grusdt2016c} differ by a factor of $8\pi$)} %(definitions in~\cite{tempere2009, grusdt2016c} differ by a factor of $8\pi$)
. That is, the impurity-boson scattering length is considerably larger than that of the boson-boson interaction.
In experiments probing the strong-coupling regime of the Bose polaron~\cite{Joergensen2016, Hu2016, yan2020bose}, this corresponds to moderate interactions.

\section{The model}

We now give a concrete model realizing the situation described above and for which we will derive an asymptotic expansion of the eigenvalues in $\rho_\mathrm{B}a_\BB^3\ll1$, featuring the Bogoliubov-Fröhlich Hamiltonian.

We consider a single impurity interacting with $N$ bosons in a box of size $L=\xi$, with periodic boundary conditions at zero temperature. That is, $N$ is proportional to $(\rho_\mathrm{B}a_\BB^3)^{-1/2}$, which is sometimes called the Gross--Pitaevskii regime. In units where $\hbar=\xi=1$, and also the impurity mass is $M=1$, the Hamiltonian is
\begin{equation*}
 \hat H = \tfrac12 \hat \bP^2 + \sum_{j=1}^N \frac{\hat \bp_j^2}{2m} + \hspace{-4pt}\sum_{1\leq i<j\leq N}\hspace{-9pt} V_\BB(\bx_i-\bx_j) + \sum_{j=1}^N  V_\IB(\bX-\bx_j),
\end{equation*}
where uppercase letters denote quantities pertaining to the impurity, lower case letters those of the bosons.
As $8\pi a_\BB/\xi=N^{-1}$, we write
\begin{equation}\label{eq:VBB}
 V_\BB(\bx)=N^2 \tilde V_\BB(N \bx),%quad  V_\IB(\bX)=N \tilde V_\IB(\sqrt N \bX).
\end{equation}
where now $\tilde a_\BB= N a_\BB=1/8\pi$ (cf. the definition~\footnote{We use here the definition of the scattering length of a general potential $4\pi a=\min \int |\nabla \phi(x)|^2 + 2m_\mathrm{red}(1+\phi(x))^2V(x) dx=2m_\mathrm{red}\int V(x)(1+\phi_\mathrm{min}(x))dx$}).
 Similarly, $a_\IB/\xi=\sqrt{\alpha} N^{-1/2}$ and we write
\begin{equation}\label{eq:VIB}
  V_\IB(\bX)=N \tilde V_\IB(\sqrt N \bX)
\end{equation}
with $\tilde a_\IB=\sqrt{N} a_\IB=\sqrt \alpha$.
We consider $\tilde V_\BB$, $\tilde V_\IB$ to be fixed while we make asymptotic expansions in $N$.

We will compare the Hamiltonian $\hat H$ with the effective Fröhlich Hamiltonian, given by
\begin{multline}
	\label{eq:HBF}
 \hat H_\mathrm{BF} = \tfrac12 \hat \bP^2 +\sum_{\bk\neq0} \omega_k b_\bk^\dagger b_{\bk} \\+ \frac{2\pi \sqrt\alpha}{\mu}\sum_{\bk\neq 0}e^{-i\bX \cdot\bk} \sqrt{\frac{k^2}{2m\omega_k}} (b_{\bk}^\dagger + b_{-\bk})
\end{multline}
where $\mu^{-1}=1+m^{-1}$ is the inverse reduced mass, $b_\bk^\dagger$, $b_{\bk}$ create and annihilate Bogoliubov phonons, whose dispersion relation is
\begin{equation*}%\label{eq:omega}
 \omega_k=\sqrt{k^2/(2m)(k^2/2m + 8\pi N a_{\BB}/m)}=\tfrac{1}{2m}\sqrt{ k^4 + 2k^2}.
\end{equation*}
Here, and in the following, sums over momenta are over $\bk\in 2\pi \mathbb{Z}/\xi$.

Note that the expression~\eqref{eq:HBF} requires renormalization~\cite{La20Bog}, so what we really mean by $\hat H_\mathrm{BF}$ is the renormalized version of~\eqref{eq:HBF} (constructed in~\cite{La20Bog,LaTr24} and explained below).

\subsection{Energy asymptotics}\label{sect:energy}

Our main results concern the asymptotic expansion of the energy for $N=\xi/(8\pi a_\BB)\gg1$ for repulsive, finite range interactions $\tilde V_\BB$, $\tilde V_\IB \geq 0$ (see \cite[Theorem 1.2]{LaTr24} for precise assumptions).
The first is an expansion of the ground state energy of $\hat H$ in units of $1/\xi^2$
\begin{align}
 E_0&=\frac{1}{4m} N  +  \frac{2\pi}{\mu}\sqrt\alpha  N^{1/2}\label{eq:E0_expand}
 \\
 &\quad -\frac{16 \pi \alpha^2}{\mu}\Big(\mu^{-1}\arcsin \mu -\sqrt{1- \mu^2} \Big) \log N + \mathcal{O}(1).\notag
 \end{align}

 In this expansion, the leading term corresponds to the usual Gross-Pitaevskii energy $4\pi a_\BB \rho_B^2 L^3$. The following term is the analogue for the boson-impurity interaction $8\pi a_\IB \rho_\mathrm{B} \rho_\mathrm{I}  L^3$. The term of order $\log N$ arises from an effective phonon-phonon interaction mediated by the impurity. It vanishes in the limit of a heavy impurity, that is, if instead of taking $M=1$ we let $M\to \infty$. Such a term has been observed in the context of the Fröhlich-Hamiltonian~\cite{grusdt2015, La20Bog}, and our study shows that it is the leading correction to the mean-field approximation. %standard contributions.
Terms that arise from similar effective three-body interactions can be found in the expansion of the energy in $a_\BB/\xi$ (without impurity)~\cite{wu1959, sawada-59, HuPi-59, caraci2023}, and in $a_\IB/\xi$~\cite{christensen2015}, but in our parameter regime these are of lower order ($N^{-1}\log N$, resp. $N^{-1/2} \log N$). This energy shift may thus be much easier to detect experimentally than the corresponding term in the pure Bose gas, which so far is only theoretical. Indeed, the shift is even larger than the Lee-Huang-Yang~\cite{LeeHuangYang-57} term, which is of order one in our units and has been observed~\cite{papp2008, navon2011, skov2021, lavoine2021}.

Our second result relates the excited eigenvalues $E_n$ to the corresponding eigenvalues $E_n^\mathrm{BF}$ of the Bogoliubov-Fr\"ohlich Hamiltonian $\hat H_\mathrm{BF}$. We have
\begin{equation}\label{eq:eigenvalues}
E_n-E_0=E_n^\mathrm{BF}-E_0^\mathrm{BF} + \mathcal{O}(N^{-\delta})
\end{equation}
for some $\delta>0$.
This shows that the excitation spectrum is well approximated by that of $\hat H_\mathrm{BF}$, and thus justifies the latter as an effective description. Our analysis also yields an approximation of the eigenstates and the dynamics, see \cite[Corollary 1.6]{LaTr24}.

% \begin{thm}\label{thm:main}
% Let $\tilde V_\BB$, $\tilde V_\IB$ be positive, have finite support and be square-integrable, and assume Bose\textendash Einstein condensation occurs. %$\tilde V_\BB$ admits Bose-Einstein condensation.
% Then, as $N= \xi/(8\pi a_\BB)\to \infty$ the ground state energy in units of $1/\xi^2$ is
% \begin{align*}
%  E_0&=\frac{1}{4m} N  +  \frac{2\pi}{\mu}\sqrt\alpha  N^{1/2}
%  \\
%  &\quad -\frac{16 \pi \alpha^2}{\mu}\Big(\mu^{-1}\arcsin \mu -\sqrt{1- \mu^2} \Big) \log N + \mathcal{O}(1).
%  \end{align*}
% Moreover, the excited eigenvalues satisfy
% \begin{equation*}
% E_n-E_0=E_n^\mathrm{BF}-E_0^\mathrm{BF} + \mathcal{O}(N^{-\delta})
% \end{equation*}
% for some $\delta>0$.
% \end{thm}

\subsection{The 2+1\textendash body problem}

The contribution to $E_0$ of order $\log N$ can be understood as a three-body effect due to the interaction of the impurity and two bosons, like the similar term for the pure Bose gas~\cite{wu1959, sawada-59, HuPi-59}. We now give a heuristic derivation of its value by considering only the 2+1\textendash body problem. An account in the $N$-body problem, and its relation to the renormalization of the Bogoliubov-Fröhlich Hamiltonian will be given later.

Ignoring the boson-boson interaction, we consider the following Hamiltonian %no direct boson-boson interaction is
\begin{equation*}
 \hat{H}_{2+1}= -\tfrac1{2\mu}(\Delta_{\br_1}+\Delta_{\br_2})+ \nabla_{\br_1}\cdot\nabla_{\br_2} + V_\IB(\br_1) +  V_\IB(\br_2),
\end{equation*}
where $\br_i$, $i=1,2$, denote the relative coordinates between each boson and the impurity.
Due to the diluteness of the system, two-body effects will dominate and we make the ansatz $\Psi(\br_1, \br_2)=f(\br_1)f(\br_2)$ for a low-energy state, with $f$ normalized, periodic.
Calculating the energy yields
\begin{equation}\label{eq:2-body min}
 \langle \Psi, \hat{H}_{2+1} \Psi \rangle=-\frac{1}{\mu}  \int \overline{f(\br)} \Delta f(\br) d^3r + 2 \int V_\IB(\br)f(\br)^2 d^3r,
\end{equation}
as the term with $\nabla_{\br_1}\cdot\nabla_{\br_2}$ integrates to zero.
The minimizer $f$ of the right hand side solves a periodized version of the zero-energy scattering equation ($f\propto 1+\phi_\IB$ from~\eqref{eq:phiIB}).
Inserting this into~\eqref{eq:2-body min} gives the leading contribution to the energy $2 E_{\text{2-bd}}\sim 4\pi N^{-1/2}\sqrt \alpha/\mu$.
In order to obtain a more precise result, we refine the ansatz by taking into account 3-body effects, $\Psi(\br_1,\br_2)=f(\br_1)f(\br_2) \Phi(\br_1,\br_2)$, where $\Phi$ differs significantly from one only where $r_1, r_2\lesssim a_\IB$.
Then, the mixed derivatives no longer integrate to zero
and play the role of an effective potential
\begin{equation}\label{eq:Veff}
V_\text{eff}(\br_1,\br_2)= \nabla f(\br_1) \cdot\nabla f(\br_2).
\end{equation}
Minimizing in $\Phi$, the three-body contribution is a generalized  scattering energy
of $V_\mathrm{eff}$~\cite{nam2023}.
Second order peturbation theory gives
\begin{align*}
 E_{\text{3-bd}}&
  \approx \sum_{\bk_1,\bk_2}  \frac{(\bk_1\cdot \bk_2)^2 |\hat f(\bk_1)|^2 |\hat f(\bk_2)|^2 }{\frac{1}{2\mu}(k_1^2+k_2^2)+  \bk_1\cdot \bk_2 }  \\
 &\approx 32\pi \alpha^2  \mu^{-1}\Big(\mu^{-1}\arcsin \mu -\sqrt{1- \mu^2} \Big) N^{-2}\log N,\notag
\end{align*}
where the second approximation replaces the sum by an integral, evaluated as in~\cite[Eq. (6.45)]{LaTr24}.
The term in the expansion of the $N$-body energy is obtained by multiplying $E_{\text{3-bd}}$ by $N^2/2$, or the number of boson pairs.

\section{Derivation of the Fr\"ohlich Hamiltonian}

To compare the $N$-body and the Bogoliubov-Fröhlich Hamiltonian we will conjugate $\hat H$ by a series of unitary transformations.

We start by rewriting the Hamiltonian in second quantization and using the isomorphism of \cite{LewNamSerSol-15}, to separate the condensate particles with $\bk=0$ from the excitations with $\bk\neq 0$ and work on the truncated Fock space of excitations $\mathcal F^{\leq N}\left(\ell^2(2\pi \mathbb{Z}^{3}/\{0\}\right))$:
\begin{subequations}
\begin{align}
 &\hat H \simeq \tfrac12 N_0(N_0-1) \hat V_\BB(0) + N_0 \hat V_\IB(0)\label{eq:HX-scalar} \\
 &\quad + \tfrac12 \hat P^2 + \frac{1}{2m}\sum_{\bk\neq 0} k^2  a_\bk^\dagger a_\bk \label{eq:HX-kin}  \\
 &\quad +N_0 \sum_{\bk\neq 0} (\hat V_\BB(k) + \hat V_\BB(0)) a_\bk^\dagger a_\bk  \label{eq:HX-diag} \\
 &\quad + \frac12\sqrt {N_0 (N_0-1)} \sum_{\bk\neq 0}  \hat V_\BB(\bk) a_\bk a_{-\bk} +\hc \label{eq:VBB-quad}\\
 &\quad +  \sqrt{N_0} \sum_{\bk\neq 0}  \hat V_\IB(\bk)  e^{-i\bX \cdot\bk}a_\bk + \hc  \label{eq:VIB-lin}  \\
 &\quad  +\sum_{\bk,\bl\neq 0} \hat V_\IB(\bk) e^{-i\bX \cdot\bk} a^\dagger_{\bk+\bl} a_\bl \label{eq:VIB-quad} \\
   &\quad + \frac12\sqrt {N_0} \sum_{\bk,\bl, \bk+\bl\neq 0}\hat V_\BB(\bk) a^\dagger_{\bk+\bl} a_\bk a_\bl +\hc  \label{eq:VBB-cube}\\
  &\quad  + \frac12\sum_{\bl+ \bk,\bmm \neq 0 \atop\bmm+\bk, \bl \neq 0}\hat V_\BB(\bk) a^\dagger_{\bl+\bk} a^\dagger_{\bmm}  a_{\bmm + \bk }a_{\bl } \label{eq:VBB-quart}
 \end{align}
\end{subequations}
where $N_0=N-\sum_{\bk\neq 0}a_\bk^\dagger a_\bk$ is the number of condensate particles.

%Our condensation hypothesis on $\tilde V_\BB$  means that on low-energy states $N_0 \geq N- c\sqrt{N}$.
Low energy states of this Hamiltonian exhibit complete Bose--Einstein condensation, $N_0\approx N$.
In view of the scaling relations (\ref{eq:VBB})\textendash (\ref{eq:VIB}), we are thus tempted to neglect the terms (\ref{eq:VIB-quad})\textendash (\ref{eq:VBB-quart}). We would then arrive at a Fröhlich-type Hamiltonian and an energy expansion as described above. %similar to Theorem~\ref{thm:main}.
However, these would feature $\hat V_{\BB}(0)$, $\hat V_{\IB}(0)$ instead of the scattering lengths $a_\BB$, $a_\IB$ and thus be incorrect even at the leading order.
Because interactions in $\hat H$ are strong on short length scales, particles correlate to balance the kinetic energy with the interaction. As a result, although most modes have a momentum of order one (that is, $\xi^{-1}$  in our units), a vanishing number of them have large momenta and are responsible for a shift in energy to leading order. We first need to factor out the short scale correlation structure, making the scattering length appear in the couplings instead of the bare potentials and changing the kinetic energy from order $N$ to order one on low energy states.

After this first step, we are left with quadratic and linear terms acting on low momenta, which up to an additive constant is essentially an ultraviolet cutoff version of
\begin{align}
\hat H_{\mathrm{Bog}} = \tfrac12 \hat P^2 + \frac{1}{2m} \sum_{\bk\neq 0} \left(k^2  + 8\pi N a_\BB\right) a^\dagger_\bk a_\bk \nonumber \\
+ \frac{1}{2m} \sum_{\bk\neq 0} 4\pi N a_\BB\left( a^\dagger_\bk a^\dagger_{-\bk} + a_\bk a_{-\bk}\right) \nonumber \\
+ \frac{1}{2\mu} \sum_{\bk\neq 0} 8 \pi \sqrt{N} a_{\IB} e^{-i\bX \cdot\bk} \left(a^\dagger_\bk + a_{-\bk}\right) \label{eq:H_LHY}
\end{align}
and which is equivalent to the Bogoliubov-Fr\"ohlich Hamiltonian (\ref{eq:HBF}) by passing to the phonon creation/annihilation operators $b^\dagger_\bk,b_\bk$, cf.~\eqref{eq:b-operators}.

Finally,  we can renormalize this operator and obtain the correct energy expansion including the $\log N$-contribution.
%Note that the Bogoliubov-Fr\"ohlich Hamiltonian does not allow for explicit diagonalization, so that the contribution of order one cannot be calculated analytically.

\subsection{Approximate diagonalization}

Large momenta are those with $k \gtrsim N^\tau$, for some small $0 <\tau < 1/2$. The unitary transformations are obtained by exponentiating linear, quadratic and cubic operators
\begin{align*}
 U_q& =\exp\Big( \tfrac12 \sum_{k>N^\tau} N\phi_\BB(\bk) a_\bk a_{-\bk}-\hc\Big), \\
  U_c& =\exp\Big(\sum_{k>N^\tau \atop \ell\leq N^\tau}\chi(N_+) \sqrt{N}\phi_\BB(\bk) a^\dagger_{\bk+\bl} a^\dagger_{-\bk} a_\bl  -\hc\Big), \\
  U_W&=\exp\Big( \sum_{k>N^\tau} \sqrt{N} \phi_\IB(\bk) e^{-i \bX\cdot\bk}a^\dagger_\bk -\hc \Big),
\end{align*}
where $N\phi_\BB$ and $N\phi_\IB$ are approximate zero-energy scattering solutions with periodic boundary conditions
\begin{align}\label{eq:phiBB}
 &\tfrac{1}{m}k^2 \phi_\BB(\bk) + \sum_{\bl\neq 0} \hat V_\BB(\bk-\bl)\phi_\BB(\bl) = -  \hat V_\BB(\bk), \\
 \label{eq:phiIB}
 &\tfrac{1}{2\mu} k^2 \phi_\IB(\bk) + \sum_{\bl\neq0} \hat V_\IB(\bk-\bl)\phi_\IB(\bl) = -  \hat V_\IB(\bk)
\end{align}
for $\bk\neq 0$, and $\chi(N_+)$ restricts the excitation number $N_+ = N - N_0$ to values much smaller than $N$.

The transformations $U_q$ and $U_c$ take care of creation and annihilation of the so-called hard and soft pairs of bosons with large momenta (the sum of which is either zero or small) and the map $U_W$ deals with the large momentum scattering between the bosons and the impurity.

From the equation (\ref{eq:phiBB}), we see that $\phi_\BB(\bk)$ behaves similarly to $-m \hat V_\BB(\bk)/k^2$, and thus, using (\ref{eq:VBB}), we obtain $\sum_{k>N^\tau} |N \phi_\BB(\bk)|^2\sim N^{-\tau}$. The unitaries are therefore close to the identity. However, since, e.g., $\sum_{k>N^\tau} k^2|N \phi_\BB(\bk)|^2\sim N$, they affect the energy to leading order. 

In order to evaluate the action of the unitaries on the Hamiltonian, we use the exact adjoint expansion
%(although the action of the linear and quadratic transformations could in principle be computed explicitly)
\begin{align*}
e^{-B} A e^{B} &= A + [A,B] + \int_0^1du \int_0^u dt e^{-tB} [[A,B],B] e^{tB} .
\end{align*}
The renormalization of the energy and the couplings becomes apparent when evaluating these commutators and simplifying the result using the scattering equations~\eqref{eq:phiBB},~\eqref{eq:phiIB}. The error terms are small relative to the kinetic operators in (\ref{eq:HX-kin}) and the positive interaction terms (\ref{eq:VIB-quad}), (\ref{eq:VBB-quart}).

For this reason we need to first implement $U_q$ to renormalize the order $N$ of the energy, followed by $U_W$ which renormalizes the order $N^{1/2}$. At this stage the kinetic energy will be of order $N^\tau$ on low energy states and we can implement $U_c$. Let us explain this in more detail.

\paragraph{First Bogoliubov transformation.}

This transformation implements the correlations associated to hard boson pairs. It acts non-trivially on the boson Hamiltonian, namely on the terms (\ref{eq:HX-kin}),(\ref{eq:VBB-quad}) and (\ref{eq:VBB-quart}). We mean by this that the other terms remain unchanged up to small errors relative to the desired precision. Using the scattering equation (\ref{eq:phiBB}) leads to the replacements
\begin{align*}
\tfrac12 N^2 \hat V_\BB(0) &\mapsto 2\pi N^2 a_\BB/m=N/(4m) \\
(\ref{eq:VBB-quad}) &\mapsto \frac1{4m} \sum_{0<k<N^\tau}   (a^\dagger_\bk a^\dagger_{-\bk} + a_\bk a_{-\bk}) - \frac{1}{2m} N_+,
\end{align*}
and extracts the energy contribution  $\sum_{0<k<N^\tau}\frac{(4\pi N a_\BB)^2}{m k^2}$ corresponding to the correction to the Born approximation which will renormalize the Bogoliubov Hamiltonian \cite{LeeHuangYang-57,BocBreCenSch-18}.  After this step, the impurity-boson interaction becomes dominant and the kinetic energy is of order $N^{1/2}$ on low energy states.

\paragraph{Weyl transformation.}\label{sec:Weyl}
The next transformation plays a similar role to $U_q$, but for $V_\IB$ instead of $V_\BB$. For fixed $\bX$, it is a Weyl transformation, mapping $U_W^\dagger a_\bk U_W=a_\bk+\sqrt N \phi_\IB(\bk)e^{-i\bX \cdot\bk}$.
The action of this unitary on the boson-boson interaction is essentially negligible, and the action on the kinetic and impurity-boson terms (\ref{eq:VIB-lin}) and (\ref{eq:VIB-quad}) has the effect of replacing
\begin{align}
N\hat V_\IB(0) &\mapsto 2\pi N a_\IB/\mu=2\pi \sqrt{\alpha N}/\mu \\
(\ref{eq:VIB-lin}) &\mapsto \frac{2\pi \sqrt\alpha}{\mu}  \sum_{0<k<N^\tau}   e^{-i\bX \cdot\bk}a^\dagger_\bk + \hc \label{eq:VIB-lin-ren}
\end{align}
and adding a correction term $\sum_{0<k<N^\tau}\frac{8 \pi^2 \alpha}{\mu k^2}$. 
The action on $\hat P^2$ also gives rise to new interaction terms, including
\begin{align}
 &\frac N2 \!\! \sum_{k,\ell>N^\tau} \!\!\!  \phi_\IB(\bk)(\bk\cdot\bl)\phi_\IB(\bl)e^{-i\bX\cdot(\bk+\bl)}
 a^\dagger_\bk a^\dagger_\bl + \hc
 \label{eq:int-tau}
\end{align}
which corresponds to the effective three-body potential (\ref{eq:Veff}) and will contribute to the $\log N$-term.

\paragraph{Cubic transformation}
Now that the kinetic energy is of order $N^\tau$ on low energy states, we can perform the cubic transformation, whose role is to take into account correlations of soft boson pairs. Acting mainly on the terms (\ref{eq:HX-kin}),(\ref{eq:VBB-cube}) and (\ref{eq:VBB-quart}), it factors out contributions from the cubic term and renormalizes the diagonal quadratic terms of the boson-boson interaction
\begin{align*}
(\ref{eq:VBB-quad}) &\mapsto \frac{1}{2m} 16\pi a_\BB N N_+  = \frac{1}{m} N_+. %\\
%(\ref{eq:VBB-cube}) &\mapsto \frac12\sqrt {N} \!\!\!\!\!\!\!\! \sum_{\{|\bk| > N^\tau, |\bl| \leq N^\tau\}^c}\hat V_\BB(\bk) a^\dagger_{\bk+\bl} a_\bk a_\bl +\hc,
\end{align*}

\subsection{Renormalization of the Fr\"ohlich Hamiltonian}

At this stage, ignoring cubic and quartic contributions, the resulting Hamiltonian is unitarily equivalent to $\hat H_{\mathrm{Bog}}$ cut off at momentum $N^{\tau}$, via a Weyl transformation similar to $U_W^\dagger$. This operator is still ultraviolet divergent when we remove the cutoff by taking $N \to \infty$ . The renormalization of the energy at order $N^\tau$ is performed again via a Bogoliubov and a Weyl transformation, which are however not close to the identity.

\paragraph{Second Bogoliubov transformation.}
To diagonalize the boson Hamiltonian, we define the usual phonon operators
\begin{equation}\label{eq:b-operators}
 b_\bk=u_k a_\bk + v_k  a^\dagger_\bk,\quad b^\dagger_\bk=u_k a^\dagger_\bk + v_k  a_\bk,
\end{equation}
with $u_k=\cosh(\theta_k)$, $v_k=\sinh(\theta_k)$ for $\theta_k= \tfrac12\log(k^2/(2m\omega_k))$ if $k<N^\tau$; otherwise $\theta_k=0$.
The Bogoliubov Hamiltonian becomes $\sum_{\bk} \omega_k b^\dagger_\bk b_\bk$ plus its ground state energy of order $N^\tau$, which combines with the scalar term created by $U_q$ to yield the usual Lee-Huang-Yang \cite{LeeHuangYang-57} correction term, which is of order one in our regime. The linear coupling between impurity and the bosons changes to
\begin{equation*}
\eqref{eq:VIB-lin-ren} \mapsto  \frac{2\pi \sqrt\alpha}{\mu}\sum_{0<k<N^\tau}  \sqrt{\frac{k^2}{2m\omega_k}} e^{-i\bX \cdot\bk}b_\bk^\dagger+\hc\,.
\end{equation*}
The transformation also creates terms of the form $\sum_{\bk,\bl} \hat V_\IB(\bk)u_{\bk+\bl}v_\bl b^\dagger_{\bk+\bl}b^\dagger_\bl$ from the quadratic interaction term~\eqref{eq:VBB-quad}, which are responsible for the logarithmic correction to the energy found in~\cite{christensen2015}. In our case this is of the lower order $N^{-1/2} \log N$, due to  the decay of $v_k\sim k^{-2}$ for large $k$ (compare~\cite[Lem.4.25]{LaTr24}).

\paragraph{Second Weyl transform.}
We can now finish regularizing the linear terms (\ref{eq:VIB-lin-ren}) with another Weyl transform, taking  into account the correct dispersion relation $\omega$ for low energy phonon-impurity scattering. Its action is given by
\begin{equation}\label{eq:Weyl2}
  b_\bk \mapsto b_\bk + e^{-i\bX\cdot\bk} \frac{2\pi \sqrt \alpha}{\mu(\tfrac12 k^2+ \omega_k)}\sqrt\frac{k^2}{2m\omega_k}.
\end{equation}
This extracts a scalar of order $N^{\tau}$ which combines with the one created by $U_W$, yielding again a contribution of order one. It also transforms the linear terms (\ref{eq:VIB-lin-ren}) to more regular quadratic terms, completing the ones in (\ref{eq:int-tau}).
Calling $U$ the product of all the unitary transformations and defining $w(\bk)$ to be equal to $2\pi\sqrt\alpha k /(\mu(\frac12 k^2+\omega_k)\sqrt{2m\omega_k})$ for $k\leq N^\tau$, and $\sqrt N \phi_\IB$ for $k>N^\tau$, we obtain
up to small errors
\begin{align}
 &U^\dagger \hat H U\approx\frac{1}{4m} N  +  \frac{2\pi}{\mu}\sqrt\alpha  N^{1/2} + E + \hat K \label{eq:BF-int-cutoff}\\
 %
 %&\quad +\tfrac12 \hat P^2 + \sum_{\bk} \omega_k b^\dagger_\bk b_\bk + \sum_{\bk,\bl\neq 0} \hat V_\IB(\bk) e^{-i\bX \cdot\bk} b^\dagger_{\bk+\bl} b_\bl \notag \\
 %
 & + \frac12 \sum_{k,\ell\neq 0} w(\bk)(\bk\cdot\bl)w(\bl)e^{-i\bX\cdot(\bk+\bl)}
b^\dagger_\bk b^\dagger_\bl +\hc +\hat R, \notag
\end{align}
where $E$ is a scalar of order one,
\begin{equation*}
 \hat K=\tfrac12 \hat P^2 + \sum_{\bk} \omega_k b^\dagger_\bk b_\bk+ \sum_{\bk,\bl\neq 0} \hat V_\IB(\bk) e^{-i\bX \cdot\bk} b^\dagger_{\bk+\bl} b_\bl , \notag\\
\end{equation*}
and $\hat R$ is the additional interaction term
\begin{multline*}
\hat R=   \sum_{k,\ell\neq 0} w(\bk)(\bk\cdot\bl)w(\bl)e^{-i\bX\cdot(\bk+\bl)}  b^\dagger_\bk b_{-\bl}  \\
 + \sum_{k\neq0 }  w(\bk) e^{-i\bX\cdot\bk} b_{\bk}^\dagger \hat P + \hc \,.
 %\label{eq:BF-int-Weyl}
\end{multline*}

\paragraph{Renormalization of $U^\dagger\hat HU$}

The Hamiltonian~\eqref{eq:BF-int-cutoff} can be renormalized non-perturbatively~\cite{La20Bog, LaTr24}.
The terms responsible for the divergence are those creating or annihilating pairs of phonons.
To deal with these, we set
\begin{align*}
\hat G=&-\frac{1}{2 \hat K} \sum_{k,\ell\neq 0} w(\bk)(\bk\cdot\bl)w(\bl)e^{-i\bX\cdot(\bk+\bl)} b^\dagger_{\bk}b^\dagger_{\bl},
\end{align*}
and rewrite the Hamiltonian as
\begin{align*}
 (1-\hat{G}^\dagger)\hat K (1-\hat G)- \hat{G}^\dagger \hat K  \hat G + \hat R,
\end{align*}
which eliminates the problematic interaction terms.
The new term $\hat{G}^\dagger \hat K G$ contains the divergent self energy.
Putting this term into normal order, we find the vacuum expectation
\begin{equation*}
 \langle \varnothing | \hat{G}^\dagger \hat K  \hat G|\varnothing\rangle \approx\frac12\sum_{\bk,\bl\neq 0}\frac{w(\bk)^2w(\bl)^2(\bk\cdot\bl)^2 }{\frac12 |\bk+\bl|^2 + \omega_{k}+\omega_\ell}
\end{equation*}
plus corrections of order one due to $\hat V_\IB(\bk)$.
Evaluating the sum for large $N$ yields the $\log N$-term from (\ref{eq:E0_expand}). 
%One easily confirms that $\langle \varnothing |\hat R^\dagger \hat K^{-1} \hat R|\varnothing\rangle$  is of order one, i.e, $\hat R$ is not singular.
The normal ordered term $\mathbin{:}\hat{G}^\dagger \hat K \hat{G}\mathbin{:}$ no longer suffers from singularities and gives a finite operator also for $N=\infty$ (since it contains an additional factor of $\hat K^{-1}$).
This is also true for $\hat R$. 
We thus have a  limiting operator for $N\to \infty$ that is unitarily equivalent to the renormalized Bogoliubov-Fr\"ohlich Hamiltonian via the inverse of the Weyl transformation~\eqref{eq:Weyl2}.
Note that the term with $\hat V_\IB(\bk)$ disappears from $\hat K$ in the limit. This term gives a global energy shift of order one, but does not change the differences of eigenvalues.

The proof of the eigenvalue approximation (\ref{eq:eigenvalues}) relies on min-max formulas, which require bounds on various error terms accumulated in the calculations.
As we have explained, these are small relative to the kinetic energies, and the positive interaction terms~\eqref{eq:VBB-quart},~\eqref{eq:VIB-quad}. They may thus be neglected for a lower bound on the energy. An upper bound is obtained by constructing trial states using the eigenstates of the Bogoliubov-Fr\"ohlich Hamiltonian and the unitary transformations.

\section{Conclusion}
We have obtained an expansion of the energy for a dilute Bose gas coupled to an impurity with effective coupling $\alpha\sim 1$. By analyzing precisely the renormalization of the model parameters due to the emergence of a point-interaction, we have shown the presence of a correction due to an effective three-body interaction. We have also established that the excitation spectrum is accurately described by the Bogoliubov-Fröhlich Hamiltonian.

\subsection*{Acknowledgments}

This work was funded by the French National Research Agency (ANR) and the German Research Foundation (DFG) through the projects (MaBoP, ANR-23-CE40-0025; PI 1114/8-1) and CRC-TRR 352 (Project-ID 470903074). J.L. acknowledges financial support by the EIPHI Graduate School (ANR-17-EURE-0002) and Bourgogne-Franche-Comté Region through the project SQC.


\begin{thebibliography}{51}%
\makeatletter
\providecommand \@ifxundefined [1]{%
 \@ifx{#1\undefined}
}%
\providecommand \@ifnum [1]{%
 \ifnum #1\expandafter \@firstoftwo
 \else \expandafter \@secondoftwo
 \fi
}%
\providecommand \@ifx [1]{%
 \ifx #1\expandafter \@firstoftwo
 \else \expandafter \@secondoftwo
 \fi
}%
\providecommand \natexlab [1]{#1}%
\providecommand \enquote  [1]{``#1''}%
\providecommand \bibnamefont  [1]{#1}%
\providecommand \bibfnamefont [1]{#1}%
\providecommand \citenamefont [1]{#1}%
\providecommand \href@noop [0]{\@secondoftwo}%
\providecommand \href [0]{\begingroup \@sanitize@url \@href}%
\providecommand \@href[1]{\@@startlink{#1}\@@href}%
\providecommand \@@href[1]{\endgroup#1\@@endlink}%
\providecommand \@sanitize@url [0]{\catcode `\\12\catcode `\$12\catcode
  `\&12\catcode `\#12\catcode `\^12\catcode `\_12\catcode `\%12\relax}%
\providecommand \@@startlink[1]{}%
\providecommand \@@endlink[0]{}%
\providecommand \url  [0]{\begingroup\@sanitize@url \@url }%
\providecommand \@url [1]{\endgroup\@href {#1}{\urlprefix }}%
\providecommand \urlprefix  [0]{URL }%
\providecommand \Eprint [0]{\href }%
\providecommand \doibase [0]{https://doi.org/}%
\providecommand \selectlanguage [0]{\@gobble}%
\providecommand \bibinfo  [0]{\@secondoftwo}%
\providecommand \bibfield  [0]{\@secondoftwo}%
\providecommand \translation [1]{[#1]}%
\providecommand \BibitemOpen [0]{}%
\providecommand \bibitemStop [0]{}%
\providecommand \bibitemNoStop [0]{.\EOS\space}%
\providecommand \EOS [0]{\spacefactor3000\relax}%
\providecommand \BibitemShut  [1]{\csname bibitem#1\endcsname}%
\let\auto@bib@innerbib\@empty
%</preamble>
\bibitem [{\citenamefont {Alexandrov}\ and\ \citenamefont
  {Devreese}(2010)}]{alexandrov2010}%
  \BibitemOpen
  \bibfield  {author} {\bibinfo {author} {\bibfnamefont {A.~S.}\ \bibnamefont
  {Alexandrov}}\ and\ \bibinfo {author} {\bibfnamefont {J.~T.}\ \bibnamefont
  {Devreese}},\ }\href@noop {} {\emph {\bibinfo {title} {Advances in polaron
  physics}}},\ \bibinfo {series} {Springer Series in Solid-State Science},
  Vol.\ \bibinfo {volume} {159}\ (\bibinfo  {publisher} {Springer},\ \bibinfo
  {year} {2010})\BibitemShut {NoStop}%
\bibitem [{\citenamefont {Grusdt}\ and\ \citenamefont
  {Demler}(2016)}]{grusdt2016c}%
  \BibitemOpen
  \bibfield  {author} {\bibinfo {author} {\bibfnamefont {F.}~\bibnamefont
  {Grusdt}}\ and\ \bibinfo {author} {\bibfnamefont {E.}~\bibnamefont
  {Demler}},\ }\bibfield  {title} {\bibinfo {title} {New theoretical approaches
  to {B}ose polarons},\ }in\ \href@noop {} {\emph {\bibinfo {booktitle}
  {Proceedings of the international school of physics "Enrico Fermi"}}},\
  \bibinfo {editor} {edited by\ \bibinfo {editor} {\bibfnamefont
  {M.}~\bibnamefont {Inguscio}}, \bibinfo {editor} {\bibfnamefont
  {W.}~\bibnamefont {Ketterle}}, \bibinfo {editor} {\bibfnamefont
  {S.}~\bibnamefont {Stringari}},\ and\ \bibinfo {editor} {\bibfnamefont
  {G.}~\bibnamefont {Roati}}}\ (\bibinfo {year} {2016})\ pp.\ \bibinfo {pages}
  {325--411}\BibitemShut {NoStop}%
\bibitem [{\citenamefont {Grusdt}\ \emph {et~al.}(2024)\citenamefont {Grusdt},
  \citenamefont {Mostaan}, \citenamefont {Demler},\ and\ \citenamefont
  {Ardila}}]{grusdt2024}%
  \BibitemOpen
  \bibfield  {author} {\bibinfo {author} {\bibfnamefont {F.}~\bibnamefont
  {Grusdt}}, \bibinfo {author} {\bibfnamefont {N.}~\bibnamefont {Mostaan}},
  \bibinfo {author} {\bibfnamefont {E.}~\bibnamefont {Demler}},\ and\ \bibinfo
  {author} {\bibfnamefont {L.~A.~P.}\ \bibnamefont {Ardila}},\ }\bibfield
  {title} {\bibinfo {title} {Impurities and polarons in bosonic quantum gases:
  a review on recent progress},\ }\href@noop {} {\bibfield  {journal} {\bibinfo
   {journal} {Rep. Prog. Phys.}\ } \textbf {\bibinfo {volume} {88}},\
  \bibinfo {pages} {066401} (\bibinfo {year} {2025})}\BibitemShut {NoStop}%
\bibitem [{\citenamefont {J\o{}rgensen}\ \emph {et~al.}(2016)\citenamefont
  {J\o{}rgensen}, \citenamefont {Wacker}, \citenamefont {Skalmstang},
  \citenamefont {Parish}, \citenamefont {Levinsen}, \citenamefont
  {Christensen}, \citenamefont {Bruun},\ and\ \citenamefont
  {Arlt}}]{Joergensen2016}%
  \BibitemOpen
  \bibfield  {author} {\bibinfo {author} {\bibfnamefont {N.~B.}\ \bibnamefont
  {J\o{}rgensen}}, \bibinfo {author} {\bibfnamefont {L.}~\bibnamefont
  {Wacker}}, \bibinfo {author} {\bibfnamefont {K.~T.}\ \bibnamefont
  {Skalmstang}}, \bibinfo {author} {\bibfnamefont {M.~M.}\ \bibnamefont
  {Parish}}, \bibinfo {author} {\bibfnamefont {J.}~\bibnamefont {Levinsen}},
  \bibinfo {author} {\bibfnamefont {R.~S.}\ \bibnamefont {Christensen}},
  \bibinfo {author} {\bibfnamefont {G.~M.}\ \bibnamefont {Bruun}},\ and\
  \bibinfo {author} {\bibfnamefont {J.~J.}\ \bibnamefont {Arlt}},\ }\bibfield
  {title} {\bibinfo {title} {Observation of attractive and repulsive polarons
  in a {B}ose-{E}instein condensate},\ }\href
  {https://doi.org/10.1103/PhysRevLett.117.055302} {\bibfield  {journal}
  {\bibinfo  {journal} {Phys. Rev. Lett.}\ }\textbf {\bibinfo {volume} {117}},\
  \bibinfo {pages} {055302} (\bibinfo {year} {2016})}\BibitemShut {NoStop}%
\bibitem [{\citenamefont {Hu}\ \emph {et~al.}(2016)\citenamefont {Hu},
  \citenamefont {Van~de Graaff}, \citenamefont {Kedar}, \citenamefont {Corson},
  \citenamefont {Cornell},\ and\ \citenamefont {Jin}}]{Hu2016}%
  \BibitemOpen
  \bibfield  {author} {\bibinfo {author} {\bibfnamefont {M.-G.}\ \bibnamefont
  {Hu}}, \bibinfo {author} {\bibfnamefont {M.~J.}\ \bibnamefont {Van~de
  Graaff}}, \bibinfo {author} {\bibfnamefont {D.}~\bibnamefont {Kedar}},
  \bibinfo {author} {\bibfnamefont {J.~P.}\ \bibnamefont {Corson}}, \bibinfo
  {author} {\bibfnamefont {E.~A.}\ \bibnamefont {Cornell}},\ and\ \bibinfo
  {author} {\bibfnamefont {D.~S.}\ \bibnamefont {Jin}},\ }\bibfield  {title}
  {\bibinfo {title} {Bose polarons in the strongly interacting regime},\ }\href
  {https://doi.org/10.1103/PhysRevLett.117.055301} {\bibfield  {journal}
  {\bibinfo  {journal} {Phys. Rev. Lett.}\ }\textbf {\bibinfo {volume} {117}},\
  \bibinfo {pages} {055301} (\bibinfo {year} {2016})}\BibitemShut {NoStop}%
\bibitem [{\citenamefont {Yan}\ \emph {et~al.}(2020)\citenamefont {Yan},
  \citenamefont {Ni}, \citenamefont {Robens},\ and\ \citenamefont
  {Zwierlein}}]{yan2020bose}%
  \BibitemOpen
  \bibfield  {author} {\bibinfo {author} {\bibfnamefont {Z.~Z.}\ \bibnamefont
  {Yan}}, \bibinfo {author} {\bibfnamefont {Y.}~\bibnamefont {Ni}}, \bibinfo
  {author} {\bibfnamefont {C.}~\bibnamefont {Robens}},\ and\ \bibinfo {author}
  {\bibfnamefont {M.~W.}\ \bibnamefont {Zwierlein}},\ }\bibfield  {title}
  {\bibinfo {title} {Bose polarons near quantum criticality},\ }\href@noop {}
  {\bibfield  {journal} {\bibinfo  {journal} {Science}\ }\textbf {\bibinfo
  {volume} {368}},\ \bibinfo {pages} {190} (\bibinfo {year}
  {2020})}\BibitemShut {NoStop}%
\bibitem [{\citenamefont {Tempere}\ \emph {et~al.}(2009)\citenamefont
  {Tempere}, \citenamefont {Casteels}, \citenamefont {Oberthaler},
  \citenamefont {Knoop}, \citenamefont {Timmermans},\ and\ \citenamefont
  {Devreese}}]{tempere2009}%
  \BibitemOpen
  \bibfield  {author} {\bibinfo {author} {\bibfnamefont {J.}~\bibnamefont
  {Tempere}}, \bibinfo {author} {\bibfnamefont {W.}~\bibnamefont {Casteels}},
  \bibinfo {author} {\bibfnamefont {M.}~\bibnamefont {Oberthaler}}, \bibinfo
  {author} {\bibfnamefont {S.}~\bibnamefont {Knoop}}, \bibinfo {author}
  {\bibfnamefont {E.}~\bibnamefont {Timmermans}},\ and\ \bibinfo {author}
  {\bibfnamefont {J.}~\bibnamefont {Devreese}},\ }\bibfield  {title} {\bibinfo
  {title} {Feynman path-integral treatment of the {BEC}-impurity polaron},\
  }\href@noop {} {\bibfield  {journal} {\bibinfo  {journal} {Phys. Rev. B}\
  }\textbf {\bibinfo {volume} {80}},\ \bibinfo {pages} {184504} (\bibinfo
  {year} {2009})}\BibitemShut {NoStop}%
\bibitem [{\citenamefont {Rath}\ and\ \citenamefont
  {Schmidt}(2013)}]{rath2013}%
  \BibitemOpen
  \bibfield  {author} {\bibinfo {author} {\bibfnamefont {S.~P.}\ \bibnamefont
  {Rath}}\ and\ \bibinfo {author} {\bibfnamefont {R.}~\bibnamefont {Schmidt}},\
  }\bibfield  {title} {\bibinfo {title} {Field-theoretical study of the {B}ose
  polaron},\ }\href@noop {} {\bibfield  {journal} {\bibinfo  {journal} {Phys.
  Rev. A.}\ }\textbf {\bibinfo {volume} {88}},\ \bibinfo {pages} {053632}
  (\bibinfo {year} {2013})}\BibitemShut {NoStop}%
\bibitem [{\citenamefont {Li}\ and\ \citenamefont {Das~Sarma}(2014)}]{li2014}%
  \BibitemOpen
  \bibfield  {author} {\bibinfo {author} {\bibfnamefont {W.}~\bibnamefont
  {Li}}\ and\ \bibinfo {author} {\bibfnamefont {S.}~\bibnamefont {Das~Sarma}},\
  }\bibfield  {title} {\bibinfo {title} {Variational study of polarons in
  {B}ose-{E}instein condensates},\ }\href@noop {} {\bibfield  {journal}
  {\bibinfo  {journal} {Phys. Rev. A}\ }\textbf {\bibinfo {volume} {90}},\
  \bibinfo {pages} {013618} (\bibinfo {year} {2014})}\BibitemShut {NoStop}%
\bibitem [{\citenamefont {Grusdt}\ \emph {et~al.}(2015)\citenamefont {Grusdt},
  \citenamefont {Shchadilova}, \citenamefont {Rubtsov},\ and\ \citenamefont
  {Demler}}]{grusdt2015}%
  \BibitemOpen
  \bibfield  {author} {\bibinfo {author} {\bibfnamefont {F.}~\bibnamefont
  {Grusdt}}, \bibinfo {author} {\bibfnamefont {Y.}~\bibnamefont {Shchadilova}},
  \bibinfo {author} {\bibfnamefont {A.}~\bibnamefont {Rubtsov}},\ and\ \bibinfo
  {author} {\bibfnamefont {E.}~\bibnamefont {Demler}},\ }\bibfield  {title}
  {\bibinfo {title} {Renormalization group approach to the {F}r{\"o}hlich
  polaron model: application to impurity-{BEC} problem},\ }\href@noop {}
  {\bibfield  {journal} {\bibinfo  {journal} {Sci. Rep.}\ }\textbf {\bibinfo
  {volume} {5}},\ \bibinfo {pages} {12124} (\bibinfo {year}
  {2015})}\BibitemShut {NoStop}%
\bibitem [{\citenamefont {Christensen}\ \emph {et~al.}(2015)\citenamefont
  {Christensen}, \citenamefont {Levinsen},\ and\ \citenamefont
  {Bruun}}]{christensen2015}%
  \BibitemOpen
  \bibfield  {author} {\bibinfo {author} {\bibfnamefont {R.}~\bibnamefont
  {Christensen}}, \bibinfo {author} {\bibfnamefont {J.}~\bibnamefont
  {Levinsen}},\ and\ \bibinfo {author} {\bibfnamefont {G.}~\bibnamefont
  {Bruun}},\ }\bibfield  {title} {\bibinfo {title} {Quasiparticle properties of
  a mobile impurity in a {B}ose-{E}instein condensate},\ }\href@noop {}
  {\bibfield  {journal} {\bibinfo  {journal} {Phys. Rev. Lett.}\ }\textbf
  {\bibinfo {volume} {115}},\ \bibinfo {pages} {160401} (\bibinfo {year}
  {2015})}\BibitemShut {NoStop}%
\bibitem [{\citenamefont {Shchadilova}\ \emph {et~al.}(2016)\citenamefont
  {Shchadilova}, \citenamefont {Schmidt}, \citenamefont {Grusdt},\ and\
  \citenamefont {Demler}}]{grusdt2016d}%
  \BibitemOpen
  \bibfield  {author} {\bibinfo {author} {\bibfnamefont {Y.}~\bibnamefont
  {Shchadilova}}, \bibinfo {author} {\bibfnamefont {R.}~\bibnamefont
  {Schmidt}}, \bibinfo {author} {\bibfnamefont {F.}~\bibnamefont {Grusdt}},\
  and\ \bibinfo {author} {\bibfnamefont {E.}~\bibnamefont {Demler}},\
  }\bibfield  {title} {\bibinfo {title} {Quantum dynamics of ultracold {B}ose
  polarons},\ }\href {https://doi.org/10.1103/PhysRevLett.117.113002}
  {\bibfield  {journal} {\bibinfo  {journal} {Phys. Rev. Lett.}\ }\textbf
  {\bibinfo {volume} {117}},\ \bibinfo {pages} {113002} (\bibinfo {year}
  {2016})}\BibitemShut {NoStop}%
\bibitem [{\citenamefont {Grusdt}\ \emph {et~al.}(2017)\citenamefont {Grusdt},
  \citenamefont {Schmidt}, \citenamefont {Shchadilova},\ and\ \citenamefont
  {Demler}}]{grusdt2017}%
  \BibitemOpen
  \bibfield  {author} {\bibinfo {author} {\bibfnamefont {F.}~\bibnamefont
  {Grusdt}}, \bibinfo {author} {\bibfnamefont {R.}~\bibnamefont {Schmidt}},
  \bibinfo {author} {\bibfnamefont {Y.~E.}\ \bibnamefont {Shchadilova}},\ and\
  \bibinfo {author} {\bibfnamefont {E.}~\bibnamefont {Demler}},\ }\bibfield
  {title} {\bibinfo {title} {Strong-coupling {B}ose polarons in a
  {B}ose-{E}instein condensate},\ }\href
  {https://doi.org/10.1103/PhysRevA.96.013607} {\bibfield  {journal} {\bibinfo
  {journal} {Phys. Rev. A}\ }\textbf {\bibinfo {volume} {96}},\ \bibinfo
  {pages} {013607} (\bibinfo {year} {2017})}\BibitemShut {NoStop}%
\bibitem [{\citenamefont {Drescher}\ \emph {et~al.}(2019)\citenamefont
  {Drescher}, \citenamefont {Salmhofer},\ and\ \citenamefont
  {Enss}}]{drescher2019}%
  \BibitemOpen
  \bibfield  {author} {\bibinfo {author} {\bibfnamefont {M.}~\bibnamefont
  {Drescher}}, \bibinfo {author} {\bibfnamefont {M.}~\bibnamefont
  {Salmhofer}},\ and\ \bibinfo {author} {\bibfnamefont {T.}~\bibnamefont
  {Enss}},\ }\bibfield  {title} {\bibinfo {title} {Real-space dynamics of
  attractive and repulsive polarons in {B}ose-{E}instein condensates},\ }\href
  {https://doi.org/10.1103/PhysRevA.99.023601} {\bibfield  {journal} {\bibinfo
  {journal} {Phys. Rev. A}\ }\textbf {\bibinfo {volume} {99}},\ \bibinfo
  {pages} {023601} (\bibinfo {year} {2019})}\BibitemShut {NoStop}%
\bibitem [{\citenamefont {Ichmoukhamedov}\ and\ \citenamefont
  {Tempere}(2019)}]{ichmoukhamedov2019}%
  \BibitemOpen
  \bibfield  {author} {\bibinfo {author} {\bibfnamefont {T.}~\bibnamefont
  {Ichmoukhamedov}}\ and\ \bibinfo {author} {\bibfnamefont {J.}~\bibnamefont
  {Tempere}},\ }\bibfield  {title} {\bibinfo {title} {Feynman path-integral
  treatment of the {B}ose polaron beyond the {F}r{\"o}hlich model},\
  }\href@noop {} {\bibfield  {journal} {\bibinfo  {journal} {Phys. Rev. A}\
  }\textbf {\bibinfo {volume} {100}},\ \bibinfo {pages} {043605} (\bibinfo
  {year} {2019})}\BibitemShut {NoStop}%
\bibitem [{\citenamefont {Levinsen}\ \emph {et~al.}(2021)\citenamefont
  {Levinsen}, \citenamefont {Ardila}, \citenamefont {Yoshida},\ and\
  \citenamefont {Parish}}]{levinsen2021}%
  \BibitemOpen
  \bibfield  {author} {\bibinfo {author} {\bibfnamefont {J.}~\bibnamefont
  {Levinsen}}, \bibinfo {author} {\bibfnamefont {L.~A.~P.}\ \bibnamefont
  {Ardila}}, \bibinfo {author} {\bibfnamefont {S.~M.}\ \bibnamefont
  {Yoshida}},\ and\ \bibinfo {author} {\bibfnamefont {M.~M.}\ \bibnamefont
  {Parish}},\ }\bibfield  {title} {\bibinfo {title} {Quantum behavior of a
  heavy impurity strongly coupled to a {B}ose gas},\ }\href@noop {} {\bibfield
  {journal} {\bibinfo  {journal} {Phys. Rev. Lett.}\ }\textbf {\bibinfo
  {volume} {127}},\ \bibinfo {pages} {033401} (\bibinfo {year}
  {2021})}\BibitemShut {NoStop}%
\bibitem [{\citenamefont {Massignan}\ \emph {et~al.}(2021)\citenamefont
  {Massignan}, \citenamefont {Yegovtsev},\ and\ \citenamefont
  {Gurarie}}]{massignan2021}%
  \BibitemOpen
  \bibfield  {author} {\bibinfo {author} {\bibfnamefont {P.}~\bibnamefont
  {Massignan}}, \bibinfo {author} {\bibfnamefont {N.}~\bibnamefont
  {Yegovtsev}},\ and\ \bibinfo {author} {\bibfnamefont {V.}~\bibnamefont
  {Gurarie}},\ }\bibfield  {title} {\bibinfo {title} {Universal aspects of a
  strongly interacting impurity in a dilute {B}ose condensate},\ }\href@noop {}
  {\bibfield  {journal} {\bibinfo  {journal} {Phys. Rev. Lett.}\ }\textbf
  {\bibinfo {volume} {126}},\ \bibinfo {pages} {123403} (\bibinfo {year}
  {2021})}\BibitemShut {NoStop}%
\bibitem [{\citenamefont {Seetharam}\ \emph {et~al.}(2021)\citenamefont
  {Seetharam}, \citenamefont {Shchadilova}, \citenamefont {Grusdt},
  \citenamefont {Zvonarev},\ and\ \citenamefont {Demler}}]{seetharam-21}%
  \BibitemOpen
  \bibfield  {author} {\bibinfo {author} {\bibfnamefont {K.}~\bibnamefont
  {Seetharam}}, \bibinfo {author} {\bibfnamefont {Y.}~\bibnamefont
  {Shchadilova}}, \bibinfo {author} {\bibfnamefont {F.}~\bibnamefont {Grusdt}},
  \bibinfo {author} {\bibfnamefont {M.~B.}\ \bibnamefont {Zvonarev}},\ and\
  \bibinfo {author} {\bibfnamefont {E.}~\bibnamefont {Demler}},\ }\bibfield
  {title} {\bibinfo {title} {Dynamical quantum {C}herenkov transition of fast
  impurities in quantum liquids},\ }\href
  {https://doi.org/10.1103/PhysRevLett.127.185302} {\bibfield  {journal}
  {\bibinfo  {journal} {Phys. Rev. Lett.}\ }\textbf {\bibinfo {volume} {127}},\
  \bibinfo {pages} {185302} (\bibinfo {year} {2021})},\ \Eprint
  {https://arxiv.org/abs/2101.00030} {2101.00030} \BibitemShut {NoStop}%
\bibitem [{\citenamefont {Ichmoukhamedov}\ and\ \citenamefont
  {Tempere}(2022)}]{ichmoukhamedov2022}%
  \BibitemOpen
  \bibfield  {author} {\bibinfo {author} {\bibfnamefont {T.}~\bibnamefont
  {Ichmoukhamedov}}\ and\ \bibinfo {author} {\bibfnamefont {J.}~\bibnamefont
  {Tempere}},\ }\bibfield  {title} {\bibinfo {title} {General memory kernels
  and further corrections to the variational path integral approach for the
  {B}ogoliubov-{F}r{\"o}hlich {H}amiltonian},\ }\href@noop {} {\bibfield
  {journal} {\bibinfo  {journal} {Phys. Rev. B}\ }\textbf {\bibinfo {volume}
  {105}},\ \bibinfo {pages} {104304} (\bibinfo {year} {2022})}\BibitemShut
  {NoStop}%
\bibitem [{\citenamefont {Mostaan}\ \emph {et~al.}(2023)\citenamefont
  {Mostaan}, \citenamefont {Goldman},\ and\ \citenamefont
  {Grusdt}}]{mostaan2023}%
  \BibitemOpen
  \bibfield  {author} {\bibinfo {author} {\bibfnamefont {N.}~\bibnamefont
  {Mostaan}}, \bibinfo {author} {\bibfnamefont {N.}~\bibnamefont {Goldman}},\
  and\ \bibinfo {author} {\bibfnamefont {F.}~\bibnamefont {Grusdt}},\
  }\bibfield  {title} {\bibinfo {title} {A unified theory of strong coupling Bose polarons:
From repulsive polarons to non-Gaussian many-body bound states},\ }\href@noop {} {\bibfield  {journal} {\bibinfo  {journal} {arXiv
  preprint arXiv:2305.00835}\ } (\bibinfo {year} {2023})}\BibitemShut {NoStop}%
\bibitem [{\citenamefont {Seetharam}\ \emph {et~al.}(2024)\citenamefont
  {Seetharam}, \citenamefont {Shchadilova}, \citenamefont {Grusdt},
  \citenamefont {Zvonarev},\ and\ \citenamefont {Demler}}]{seetharam-24}%
  \BibitemOpen
  \bibfield  {author} {\bibinfo {author} {\bibfnamefont {K.}~\bibnamefont
  {Seetharam}}, \bibinfo {author} {\bibfnamefont {Y.}~\bibnamefont
  {Shchadilova}}, \bibinfo {author} {\bibfnamefont {F.}~\bibnamefont {Grusdt}},
  \bibinfo {author} {\bibfnamefont {M.}~\bibnamefont {Zvonarev}},\ and\
  \bibinfo {author} {\bibfnamefont {E.}~\bibnamefont {Demler}},\ }\bibfield
  {title} {\bibinfo {title} {Quantum {C}herenkov transition of finite-momentum
  {B}ose polarons},\ }\href {https://doi.org/10.1103/PhysRevA.110.063306}
  {\bibfield  {journal} {\bibinfo  {journal} {Phys. Rev. A}\ }\textbf {\bibinfo
  {volume} {110}},\ \bibinfo {pages} {063306} (\bibinfo {year}
  {2024})}\BibitemShut {NoStop}%
\bibitem [{\citenamefont {Rodnianski}\ and\ \citenamefont
  {Schlein}(2009)}]{RodSch-09}%
  \BibitemOpen
  \bibfield  {author} {\bibinfo {author} {\bibfnamefont {I.}~\bibnamefont
  {Rodnianski}}\ and\ \bibinfo {author} {\bibfnamefont {B.}~\bibnamefont
  {Schlein}},\ }\bibfield  {title} {\bibinfo {title} {Quantum fluctuations and
  rate of convergence towards mean field dynamics},\ }\href
  {https://doi.org/10.1007/s00220-009-0867-4} {\bibfield  {journal} {\bibinfo
  {journal} {Commun. Math. Phys.}\ }\textbf {\bibinfo {volume} {291}},\
  \bibinfo {pages} {31} (\bibinfo {year} {2009})}\BibitemShut {NoStop}%
\bibitem [{\citenamefont {Yau}\ and\ \citenamefont {Yin}(2009)}]{YauYin-09}%
  \BibitemOpen
  \bibfield  {author} {\bibinfo {author} {\bibfnamefont {H.-T.}\ \bibnamefont
  {Yau}}\ and\ \bibinfo {author} {\bibfnamefont {J.}~\bibnamefont {Yin}},\
  }\bibfield  {title} {\bibinfo {title} {The second order upper bound for the
  ground energy of a {B}ose gas},\ }\href
  {https://doi.org/10.1007/s10955-009-9792-3} {\bibfield  {journal} {\bibinfo
  {journal} {J. Stat. Phys.}\ }\textbf {\bibinfo {volume} {136}},\ \bibinfo
  {pages} {453} (\bibinfo {year} {2009})}\BibitemShut {NoStop}%
\bibitem [{\citenamefont {Grillakis}\ \emph {et~al.}(2010)\citenamefont
  {Grillakis}, \citenamefont {Machedon},\ and\ \citenamefont
  {Margetis}}]{GriMacMar-10}%
  \BibitemOpen
  \bibfield  {author} {\bibinfo {author} {\bibfnamefont {M.~G.}\ \bibnamefont
  {Grillakis}}, \bibinfo {author} {\bibfnamefont {M.}~\bibnamefont
  {Machedon}},\ and\ \bibinfo {author} {\bibfnamefont {D.}~\bibnamefont
  {Margetis}},\ }\bibfield  {title} {\bibinfo {title} {Second-order corrections
  to mean field evolution of weakly interacting bosons. {I}},\ }\href
  {https://doi.org/10.1007/s00220-009-0933-y} {\bibfield  {journal} {\bibinfo
  {journal} {Commun. Math. Phys.}\ }\textbf {\bibinfo {volume} {294}},\
  \bibinfo {pages} {273} (\bibinfo {year} {2010})}\BibitemShut {NoStop}%
\bibitem [{\citenamefont {Grillakis}\ \emph {et~al.}(2011)\citenamefont
  {Grillakis}, \citenamefont {Machedon},\ and\ \citenamefont
  {Margetis}}]{GriMacMar-11}%
  \BibitemOpen
  \bibfield  {author} {\bibinfo {author} {\bibfnamefont {M.~G.}\ \bibnamefont
  {Grillakis}}, \bibinfo {author} {\bibfnamefont {M.}~\bibnamefont
  {Machedon}},\ and\ \bibinfo {author} {\bibfnamefont {D.}~\bibnamefont
  {Margetis}},\ }\bibfield  {title} {\bibinfo {title} {Second-order corrections
  to mean field evolution of weakly interacting bosons. {II}},\ }\href
  {https://doi.org/10.1016/j.aim.2011.06.028} {\bibfield  {journal} {\bibinfo
  {journal} {Adv. Math.}\ }\textbf {\bibinfo {volume} {228}},\ \bibinfo {pages}
  {1788} (\bibinfo {year} {2011})}\BibitemShut {NoStop}%
\bibitem [{\citenamefont {Seiringer}(2011)}]{Sei-11}%
  \BibitemOpen
  \bibfield  {author} {\bibinfo {author} {\bibfnamefont {R.}~\bibnamefont
  {Seiringer}},\ }\bibfield  {title} {\bibinfo {title} {The excitation spectrum
  for weakly interacting bosons},\ }\href
  {https://doi.org/10.1007/s00220-011-1261-6} {\bibfield  {journal} {\bibinfo
  {journal} {Comm. Math. Phys.}\ }\textbf {\bibinfo {volume} {306}},\ \bibinfo
  {pages} {565} (\bibinfo {year} {2011})}\BibitemShut {NoStop}%
\bibitem [{\citenamefont {Lewin}\ \emph
  {et~al.}(2015{\natexlab{a}})\citenamefont {Lewin}, \citenamefont {Nam},\ and\
  \citenamefont {Schlein}}]{LewNamSch-15}%
  \BibitemOpen
  \bibfield  {author} {\bibinfo {author} {\bibfnamefont {M.}~\bibnamefont
  {Lewin}}, \bibinfo {author} {\bibfnamefont {P.~T.}\ \bibnamefont {Nam}},\
  and\ \bibinfo {author} {\bibfnamefont {B.}~\bibnamefont {Schlein}},\
  }\bibfield  {title} {\bibinfo {title} {Fluctuations around {H}artree states
  in the mean-field regime},\ }\href {https://doi.org/10.1353/ajm.2015.0040}
  {\bibfield  {journal} {\bibinfo  {journal} {Amer. J. Math.}\ }\textbf
  {\bibinfo {volume} {137}},\ \bibinfo {pages} {1613} (\bibinfo {year}
  {2015}{\natexlab{a}})}\BibitemShut {NoStop}%
\bibitem [{\citenamefont {Lewin}\ \emph
  {et~al.}(2015{\natexlab{b}})\citenamefont {Lewin}, \citenamefont {Nam},
  \citenamefont {Serfaty},\ and\ \citenamefont {Solovej}}]{LewNamSerSol-15}%
  \BibitemOpen
  \bibfield  {author} {\bibinfo {author} {\bibfnamefont {M.}~\bibnamefont
  {Lewin}}, \bibinfo {author} {\bibfnamefont {P.~T.}\ \bibnamefont {Nam}},
  \bibinfo {author} {\bibfnamefont {S.}~\bibnamefont {Serfaty}},\ and\ \bibinfo
  {author} {\bibfnamefont {J.~P.}\ \bibnamefont {Solovej}},\ }\bibfield
  {title} {\bibinfo {title} {Bogoliubov spectrum of interacting {B}ose gases},\
  }\href@noop {} {\bibfield  {journal} {\bibinfo  {journal} {Comm. Pure Appl.
  Math.}\ }\textbf {\bibinfo {volume} {68}},\ \bibinfo {pages} {413} (\bibinfo
  {year} {2015}{\natexlab{b}})},\ \Eprint {https://arxiv.org/abs/1211.2778}
  {1211.2778} \BibitemShut {NoStop}%
\bibitem [{\citenamefont {Boccato}\ \emph {et~al.}(2017)\citenamefont
  {Boccato}, \citenamefont {Cenatiempo},\ and\ \citenamefont
  {Schlein}}]{BocCenSch-17}%
  \BibitemOpen
  \bibfield  {author} {\bibinfo {author} {\bibfnamefont {C.}~\bibnamefont
  {Boccato}}, \bibinfo {author} {\bibfnamefont {S.}~\bibnamefont
  {Cenatiempo}},\ and\ \bibinfo {author} {\bibfnamefont {B.}~\bibnamefont
  {Schlein}},\ }\bibfield  {title} {\bibinfo {title} {Quantum many-body
  fluctuations around nonlinear {S}chr\"{o}dinger dynamics},\ }\href
  {https://doi.org/10.1007/s00023-016-0513-6} {\bibfield  {journal} {\bibinfo
  {journal} {Ann. Henri Poincar\'{e}}\ }\textbf {\bibinfo {volume} {18}},\
  \bibinfo {pages} {113} (\bibinfo {year} {2017})}\BibitemShut {NoStop}%
\bibitem [{\citenamefont {Boccato}\ \emph {et~al.}(2019)\citenamefont
  {Boccato}, \citenamefont {Brennecke}, \citenamefont {Cenatiempo},\ and\
  \citenamefont {Schlein}}]{BocBreCenSch-18}%
  \BibitemOpen
  \bibfield  {author} {\bibinfo {author} {\bibfnamefont {C.}~\bibnamefont
  {Boccato}}, \bibinfo {author} {\bibfnamefont {C.}~\bibnamefont {Brennecke}},
  \bibinfo {author} {\bibfnamefont {S.}~\bibnamefont {Cenatiempo}},\ and\
  \bibinfo {author} {\bibfnamefont {B.}~\bibnamefont {Schlein}},\ }\bibfield
  {title} {\bibinfo {title} {Bogoliubov theory in the {G}ross-{P}itaevskii
  limit},\ }\href {https://doi.org/10.4310/ACTA.2019.v222.n2.a1} {\bibfield
  {journal} {\bibinfo  {journal} {Acta Math.}\ }\textbf {\bibinfo {volume}
  {222}},\ \bibinfo {pages} {219} (\bibinfo {year} {2019})}\BibitemShut
  {NoStop}%
\bibitem [{\citenamefont {Fournais}\ and\ \citenamefont
  {Solovej}(2020)}]{FouSol-20}%
  \BibitemOpen
  \bibfield  {author} {\bibinfo {author} {\bibfnamefont {S.}~\bibnamefont
  {Fournais}}\ and\ \bibinfo {author} {\bibfnamefont {J.~P.}\ \bibnamefont
  {Solovej}},\ }\bibfield  {title} {\bibinfo {title} {The energy of dilute
  {B}ose gases},\ }\href@noop {} {\bibfield  {journal} {\bibinfo  {journal}
  {Ann. Math. (2)}\ }\textbf {\bibinfo {volume} {192}},\ \bibinfo {pages} {893}
  (\bibinfo {year} {2020})}\BibitemShut {NoStop}%
\bibitem [{\citenamefont {Fournais}\ and\ \citenamefont
  {Solovej}(2023)}]{FouSol-23}%
  \BibitemOpen
  \bibfield  {author} {\bibinfo {author} {\bibfnamefont {S.}~\bibnamefont
  {Fournais}}\ and\ \bibinfo {author} {\bibfnamefont {J.~P.}\ \bibnamefont
  {Solovej}},\ }\bibfield  {title} {\bibinfo {title} {The energy of dilute
  {B}ose gases {II}: The general case},\ }\href@noop {} {\bibfield  {journal}
  {\bibinfo  {journal} {Invent. Math.}\ }\textbf {\bibinfo {volume} {232}},\
  \bibinfo {pages} {863} (\bibinfo {year} {2023})}\BibitemShut {NoStop}%
\bibitem [{\citenamefont {Nam}\ and\ \citenamefont {Triay}(2023)}]{NamTri-23}%
  \BibitemOpen
  \bibfield  {author} {\bibinfo {author} {\bibfnamefont {P.~T.}\ \bibnamefont
  {Nam}}\ and\ \bibinfo {author} {\bibfnamefont {A.}~\bibnamefont {Triay}},\
  }\bibfield  {title} {\bibinfo {title} {{B}ogoliubov excitation spectrum of
  trapped {B}ose gases in the {G}ross-{P}itaevskii regime},\ }\href
  {https://doi.org/10.1016/j.matpur.2023.06.002} {\bibfield  {journal}
  {\bibinfo  {journal} {J. Math. Pures Appl. (9)}\ }\textbf {\bibinfo {volume}
  {176}},\ \bibinfo {pages} {18} (\bibinfo {year} {2023})}\BibitemShut
  {NoStop}%
\bibitem [{\citenamefont {Brennecke}\ \emph {et~al.}(2022)\citenamefont
  {Brennecke}, \citenamefont {Schlein},\ and\ \citenamefont
  {Schraven}}]{BreSchSch-22}%
  \BibitemOpen
  \bibfield  {author} {\bibinfo {author} {\bibfnamefont {C.}~\bibnamefont
  {Brennecke}}, \bibinfo {author} {\bibfnamefont {B.}~\bibnamefont {Schlein}},\
  and\ \bibinfo {author} {\bibfnamefont {S.}~\bibnamefont {Schraven}},\
  }\bibfield  {title} {\bibinfo {title} {{B}ogoliubov theory for trapped bosons
  in the {G}ross-{P}itaevskii regime},\ }\href
  {https://doi.org/10.1007/s00023-021-01151-z} {\bibfield  {journal} {\bibinfo
  {journal} {Ann. H. Poincar\'{e}}\ }\textbf {\bibinfo {volume} {23}},\
  \bibinfo {pages} {1583} (\bibinfo {year} {2022})}\BibitemShut {NoStop}%
\bibitem [{\citenamefont {Haberberger}\ \emph {et~al.}(2023)\citenamefont
  {Haberberger}, \citenamefont {Hainzl}, \citenamefont {Nam}, \citenamefont
  {Seiringer},\ and\ \citenamefont {Triay}}]{HabHaiNamSeiTri-23}%
  \BibitemOpen
  \bibfield  {author} {\bibinfo {author} {\bibfnamefont {F.}~\bibnamefont
  {Haberberger}}, \bibinfo {author} {\bibfnamefont {C.}~\bibnamefont {Hainzl}},
  \bibinfo {author} {\bibfnamefont {P.~T.}\ \bibnamefont {Nam}}, \bibinfo
  {author} {\bibfnamefont {R.}~\bibnamefont {Seiringer}},\ and\ \bibinfo
  {author} {\bibfnamefont {A.}~\bibnamefont {Triay}},\ }\bibfield  {title}
  {\bibinfo {title} {The free energy of dilute {B}ose gases at low
  temperatures},\ }\href@noop {} {\bibfield  {journal} {\bibinfo  {journal}
  {arXiv preprint arXiv:2304.02405}\ } (\bibinfo {year} {2023})}\BibitemShut
  {NoStop}%
\bibitem [{\citenamefont {Caraci}\ \emph {et~al.}(2024)\citenamefont {Caraci},
  \citenamefont {Oldenburg},\ and\ \citenamefont {Schlein}}]{caraci-24}%
  \BibitemOpen
  \bibfield  {author} {\bibinfo {author} {\bibfnamefont {C.}~\bibnamefont
  {Caraci}}, \bibinfo {author} {\bibfnamefont {J.}~\bibnamefont {Oldenburg}},\
  and\ \bibinfo {author} {\bibfnamefont {B.}~\bibnamefont {Schlein}},\
  }\bibfield  {title} {\bibinfo {title} {Quantum fluctuations of many-body
  dynamics around the {G}ross--{P}itaevskii equation},\ }\href@noop {}
  {\bibfield  {journal} {\bibinfo  {journal} {Ann. Inst. Henri Poincar{\'e} C}\
  } (\bibinfo {year} {2024})}\BibitemShut {NoStop}%
\bibitem [{\citenamefont {My{\'s}liwy}\ and\ \citenamefont
  {Seiringer}(2020)}]{MySe-20}%
  \BibitemOpen
  \bibfield  {author} {\bibinfo {author} {\bibfnamefont {K.}~\bibnamefont
  {My{\'s}liwy}}\ and\ \bibinfo {author} {\bibfnamefont {R.}~\bibnamefont
  {Seiringer}},\ }\bibfield  {title} {\bibinfo {title} {Microscopic derivation
  of the {F}r{\"o}hlich {H}amiltonian for the {B}ose polaron in the mean-field
  limit},\ }\href@noop {} {\bibfield  {journal} {\bibinfo  {journal} {Ann. H.
  Poincar{\'e}}\ }\textbf {\bibinfo {volume} {21}},\ \bibinfo {pages} {4003}
  (\bibinfo {year} {2020})}\BibitemShut {NoStop}%
\bibitem [{\citenamefont {Lampart}\ and\ \citenamefont {Pickl}(2022)}]{LaPi22}%
  \BibitemOpen
  \bibfield  {author} {\bibinfo {author} {\bibfnamefont {J.}~\bibnamefont
  {Lampart}}\ and\ \bibinfo {author} {\bibfnamefont {P.}~\bibnamefont
  {Pickl}},\ }\bibfield  {title} {\bibinfo {title} {Dynamics of a tracer
  particle interacting with excitations of a {B}ose--{E}instein condensate},\
  }\href@noop {} {\bibfield  {journal} {\bibinfo  {journal} {Ann. H.
  Poincar{\'e}}\ }\textbf {\bibinfo {volume} {23}},\ \bibinfo {pages} {2855}
  (\bibinfo {year} {2022})}\BibitemShut {NoStop}%
\bibitem [{\citenamefont {Lampart}\ and\ \citenamefont {Triay}(2024)}]{LaTr24}%
  \BibitemOpen
  \bibfield  {author} {\bibinfo {author} {\bibfnamefont {J.}~\bibnamefont
  {Lampart}}\ and\ \bibinfo {author} {\bibfnamefont {A.}~\bibnamefont
  {Triay}},\ }\bibfield  {title} {\bibinfo {title} {The excitation spectrum of
  a {B}ose gas with an impurity in the {G}ross--{P}itaevskii regime},\
  }\href@noop {}
  \href
  {https://doi.org/10.1007/s00205-025-02112-0} {\bibfield  {journal} {\bibinfo  {journal} { Arch. Rational Mech. Anal.}\ }\textbf {\bibinfo
  {volume} {249}},\ (\bibinfo {year}
  {2025})}\BibitemShut {NoStop}%
\bibitem [{\citenamefont {Wu}(1959)}]{wu1959}%
  \BibitemOpen
  \bibfield  {author} {\bibinfo {author} {\bibfnamefont {T.~T.}\ \bibnamefont
  {Wu}},\ }\bibfield  {title} {\bibinfo {title} {Ground state of a {B}ose
  system of hard spheres},\ }\href@noop {} {\bibfield  {journal} {\bibinfo
  {journal} {Phys. Rev.}\ }\textbf {\bibinfo {volume} {115}},\ \bibinfo {pages}
  {1390} (\bibinfo {year} {1959})}\BibitemShut {NoStop}%
\bibitem [{\citenamefont {Sawada}(1959)}]{sawada-59}%
  \BibitemOpen
  \bibfield  {author} {\bibinfo {author} {\bibfnamefont {K.}~\bibnamefont
  {Sawada}},\ }\bibfield  {title} {\bibinfo {title} {Ground-state energy of
  {B}ose-{E}instein gas with repulsive interaction},\ }\href@noop {} {\bibfield
   {journal} {\bibinfo  {journal} {Phys. Rev.}\ }\textbf {\bibinfo {volume}
  {116}},\ \bibinfo {pages} {1344} (\bibinfo {year} {1959})}\BibitemShut
  {NoStop}%
\bibitem [{\citenamefont {Hugenholtz}\ and\ \citenamefont
  {Pines}(1959)}]{HuPi-59}%
  \BibitemOpen
  \bibfield  {author} {\bibinfo {author} {\bibfnamefont {N.}~\bibnamefont
  {Hugenholtz}}\ and\ \bibinfo {author} {\bibfnamefont {D.}~\bibnamefont
  {Pines}},\ }\bibfield  {title} {\bibinfo {title} {Ground-state energy and
  excitation spectrum of a system of interacting bosons},\ }\href@noop {}
  {\bibfield  {journal} {\bibinfo  {journal} {Phys. Rev.}\ }\textbf {\bibinfo
  {volume} {116}},\ \bibinfo {pages} {489} (\bibinfo {year}
  {1959})}\BibitemShut {NoStop}%
\bibitem [{\citenamefont {Caraci}\ \emph {et~al.}(2023)\citenamefont {Caraci},
  \citenamefont {Olgiati}, \citenamefont {Aubin},\ and\ \citenamefont
  {Schlein}}]{caraci2023}%
  \BibitemOpen
  \bibfield  {author} {\bibinfo {author} {\bibfnamefont {C.}~\bibnamefont
  {Caraci}}, \bibinfo {author} {\bibfnamefont {A.}~\bibnamefont {Olgiati}},
  \bibinfo {author} {\bibfnamefont {D.~S.}\ \bibnamefont {Aubin}},\ and\
  \bibinfo {author} {\bibfnamefont {B.}~\bibnamefont {Schlein}},\ }\bibfield
  {title} {\bibinfo {title} {Third order corrections to the ground state energy
  of a {B}ose gas in the {G}ross-{P}itaevskii regime},\ }\href@noop {}
  {\bibfield  {journal} {\bibinfo  {journal} {Commun. Math. Phys.}\
  }\textbf{\bibinfo {volume} {406}} (\bibinfo {year} {2025})}\BibitemShut {NoStop}%
\bibitem [{Note1()}]{Note1}%
  \BibitemOpen
  \bibinfo {note} {Note that definitions of this parameter in~\cite {tempere2009, grusdt2016c} differ by a factor of $8\pi $)}\BibitemShut {NoStop}%
  \bibitem [{Note2()}]{Note2}%
  \BibitemOpen
  \bibinfo {note} {We use here the definition of the scattering length of a
  general potential $4\pi a=\protect \qopname \relax m{min}\DOTSI \intop
  \ilimits@ |\nabla \varphi (x)|^2 + 2m_\protect \mathrm {red}(1+\varphi
  (x))^2V(x) dx=2m_\protect \mathrm {red}\DOTSI \intop \ilimits@ V(x)(1+\varphi
  _\protect \mathrm {min}(x))dx$}\BibitemShut {NoStop}%
\bibitem [{\citenamefont {Lampart}(2020)}]{La20Bog}%
  \BibitemOpen
  \bibfield  {author} {\bibinfo {author} {\bibfnamefont {J.}~\bibnamefont
  {Lampart}},\ }\bibfield  {title} {\bibinfo {title} {The renormalised
  {B}ogoliubov-{F}r{\"o}hlich {H}amiltonian},\ }\href@noop {} {\bibfield
  {journal} {\bibinfo  {journal} {J. Math. Phys.}\ }\textbf {\bibinfo {volume}
  {61}},\ \bibinfo {pages} {101902} (\bibinfo {year} {2020})}\BibitemShut
  {NoStop}%
\bibitem [{\citenamefont {Lee}\ \emph {et~al.}(1957)\citenamefont {Lee},
  \citenamefont {Huang},\ and\ \citenamefont {Yang}}]{LeeHuangYang-57}%
  \BibitemOpen
  \bibfield  {author} {\bibinfo {author} {\bibfnamefont {T.}~\bibnamefont
  {Lee}}, \bibinfo {author} {\bibfnamefont {K.}~\bibnamefont {Huang}},\ and\
  \bibinfo {author} {\bibfnamefont {C.}~\bibnamefont {Yang}},\ }\bibfield
  {title} {\bibinfo {title} {{Eigenvalues and eigenfunctions of a {B}ose system
  of hard spheres and its low-temperature properties}},\ }\href@noop {}
  {\bibfield  {journal} {\bibinfo  {journal} {Phys. Rev.}\ }\textbf {\bibinfo
  {volume} {106}},\ \bibinfo {pages} {1135} (\bibinfo {year}
  {1957})}\BibitemShut {NoStop}%
\bibitem [{\citenamefont {Papp}\ \emph {et~al.}(2008)\citenamefont {Papp},
  \citenamefont {Pino}, \citenamefont {Wild}, \citenamefont {Ronen},
  \citenamefont {Wieman}, \citenamefont {Jin},\ and\ \citenamefont
  {Cornell}}]{papp2008}%
  \BibitemOpen
  \bibfield  {author} {\bibinfo {author} {\bibfnamefont {S.}~\bibnamefont
  {Papp}}, \bibinfo {author} {\bibfnamefont {J.}~\bibnamefont {Pino}}, \bibinfo
  {author} {\bibfnamefont {R.}~\bibnamefont {Wild}}, \bibinfo {author}
  {\bibfnamefont {S.}~\bibnamefont {Ronen}}, \bibinfo {author} {\bibfnamefont
  {C.~E.}\ \bibnamefont {Wieman}}, \bibinfo {author} {\bibfnamefont {D.~S.}\
  \bibnamefont {Jin}},\ and\ \bibinfo {author} {\bibfnamefont {E.~A.}\
  \bibnamefont {Cornell}},\ }\bibfield  {title} {\bibinfo {title} {Bragg
  spectroscopy of a strongly interacting {Rb85} {B}ose-{E}instein condensate},\
  }\href@noop {} {\bibfield  {journal} {\bibinfo  {journal} {Phys. Rev. Lett.}\
  }\textbf {\bibinfo {volume} {101}},\ \bibinfo {pages} {135301} (\bibinfo
  {year} {2008})}\BibitemShut {NoStop}%
\bibitem [{\citenamefont {Navon}\ \emph {et~al.}(2011)\citenamefont {Navon},
  \citenamefont {Piatecki}, \citenamefont {G{\"u}nter}, \citenamefont {Rem},
  \citenamefont {Nguyen}, \citenamefont {Chevy}, \citenamefont {Krauth},\ and\
  \citenamefont {Salomon}}]{navon2011}%
  \BibitemOpen
  \bibfield  {author} {\bibinfo {author} {\bibfnamefont {N.}~\bibnamefont
  {Navon}}, \bibinfo {author} {\bibfnamefont {S.}~\bibnamefont {Piatecki}},
  \bibinfo {author} {\bibfnamefont {K.}~\bibnamefont {G{\"u}nter}}, \bibinfo
  {author} {\bibfnamefont {B.}~\bibnamefont {Rem}}, \bibinfo {author}
  {\bibfnamefont {T.~C.}\ \bibnamefont {Nguyen}}, \bibinfo {author}
  {\bibfnamefont {F.}~\bibnamefont {Chevy}}, \bibinfo {author} {\bibfnamefont
  {W.}~\bibnamefont {Krauth}},\ and\ \bibinfo {author} {\bibfnamefont
  {C.}~\bibnamefont {Salomon}},\ }\bibfield  {title} {\bibinfo {title}
  {Dynamics and thermodynamics of the low-temperature strongly interacting
  {B}ose gas},\ }\href@noop {} {\bibfield  {journal} {\bibinfo  {journal}
  {Phys. Rev. Lett.}\ }\textbf {\bibinfo {volume} {107}},\ \bibinfo {pages}
  {135301} (\bibinfo {year} {2011})}\BibitemShut {NoStop}%
\bibitem [{\citenamefont {Skov}\ \emph {et~al.}(2021)\citenamefont {Skov},
  \citenamefont {Skou}, \citenamefont {J{\o}rgensen},\ and\ \citenamefont
  {Arlt}}]{skov2021}%
  \BibitemOpen
  \bibfield  {author} {\bibinfo {author} {\bibfnamefont {T.~G.}\ \bibnamefont
  {Skov}}, \bibinfo {author} {\bibfnamefont {M.~G.}\ \bibnamefont {Skou}},
  \bibinfo {author} {\bibfnamefont {N.~B.}\ \bibnamefont {J{\o}rgensen}},\ and\
  \bibinfo {author} {\bibfnamefont {J.~J.}\ \bibnamefont {Arlt}},\ }\bibfield
  {title} {\bibinfo {title} {Observation of a {L}ee-{H}uang-{Y}ang fluid},\
  }\href@noop {} {\bibfield  {journal} {\bibinfo  {journal} {Phys. Rev. Lett.}\
  }\textbf {\bibinfo {volume} {126}},\ \bibinfo {pages} {230404} (\bibinfo
  {year} {2021})}\BibitemShut {NoStop}%
\bibitem [{\citenamefont {Lavoine}\ \emph {et~al.}(2021)\citenamefont
  {Lavoine}, \citenamefont {Hammond}, \citenamefont {Recati}, \citenamefont
  {Petrov},\ and\ \citenamefont {Bourdel}}]{lavoine2021}%
  \BibitemOpen
  \bibfield  {author} {\bibinfo {author} {\bibfnamefont {L.}~\bibnamefont
  {Lavoine}}, \bibinfo {author} {\bibfnamefont {A.}~\bibnamefont {Hammond}},
  \bibinfo {author} {\bibfnamefont {A.}~\bibnamefont {Recati}}, \bibinfo
  {author} {\bibfnamefont {D.}~\bibnamefont {Petrov}},\ and\ \bibinfo {author}
  {\bibfnamefont {T.}~\bibnamefont {Bourdel}},\ }\bibfield  {title} {\bibinfo
  {title} {Beyond-mean-field effects in {R}abi-coupled two-component
  {B}ose-{E}instein condensate},\ }\href@noop {} {\bibfield  {journal}
  {\bibinfo  {journal} {Phys. Rev. Lett.}\ }\textbf {\bibinfo {volume} {127}},\
  \bibinfo {pages} {203402} (\bibinfo {year} {2021})}\BibitemShut {NoStop}%
\bibitem [{\citenamefont {Nam}\ \emph {et~al.}(2023)\citenamefont {Nam},
  \citenamefont {Ricaud},\ and\ \citenamefont {Triay}}]{nam2023}%
  \BibitemOpen
  \bibfield  {author} {\bibinfo {author} {\bibfnamefont {P.~T.}\ \bibnamefont
  {Nam}}, \bibinfo {author} {\bibfnamefont {J.}~\bibnamefont {Ricaud}},\ and\
  \bibinfo {author} {\bibfnamefont {A.}~\bibnamefont {Triay}},\ }\bibfield
  {title} {\bibinfo {title} {The condensation of a trapped dilute {B}ose gas
  with three-body interactions},\ }\href@noop {} {\bibfield  {journal}
  {\bibinfo  {journal} {Prob. Math. Phys.}\ }\textbf {\bibinfo {volume} {4}},\
  \bibinfo {pages} {91} (\bibinfo {year} {2023})}\BibitemShut {NoStop}%
\end{thebibliography}
\end{document}